\documentclass[conference]{IEEEtran}
\IEEEoverridecommandlockouts
\usepackage[left=1.5cm,right=1.5cm,top=1.7cm]{geometry}
\usepackage{cite}
\usepackage{subcaption}
\usepackage{graphicx}
\usepackage{lipsum}
\usepackage[symbol]{footmisc}
\usepackage{amsmath,amssymb,amsfonts}
\usepackage{graphicx}
\usepackage{textcomp}
\usepackage{xcolor}
\usepackage[linesnumbered,commentsnumbered,ruled,vlined]{algorithm2e}
\usepackage{caption}
\usepackage{algorithmic}
\usepackage[ruled,vlined]{algorithm2e}
\usepackage{amsmath}

\def\BibTeX{{\rm B\kern-.05em{\sc i\kern-.025em b}\kern-.08em
    T\kern-.1667em\lower.7ex\hbox{E}\kern-.125emX}}
\begin{document}
\renewcommand{\thefootnote}{\fnsymbol{footnote}}

% \footnote[num]{text}
% \title{FlowPath: A Flow level Inspection Tool for Identifying and Addressing Network Path Imbalance in AI Workloads\\
\title{FlowTracer: A Tool for Uncovering Network Path Usage Imbalance in AI Training Clusters\\
}
% \author{
%     Hasibul Jamil\textsuperscript{1}, Abdul Alim\textsuperscript{2}, Laurent Schares\textsuperscript{2}, Pavlos Maniotis\textsuperscript{2},\\
%     Tevfik Kosar\textsuperscript{1}, Ali Sydney\textsuperscript{2}, Liran Schour\textsuperscript{3}, Bengi Karacali-Akyamac\textsuperscript{2}\\
%     \\
%     \textsuperscript{1}University at Buffalo (SUNY), Buffalo, New York, USA\\
%     Email: \{mdhasibu, tkosar\}@buffalo.edu\\
%     \\
%     \textsuperscript{2}IBM Research, Yorktown Heights, New York, USA\\
%     Email: \{malim, schares, ppmaniotis, sydney, bkaracali\}@us.ibm.com\\
%     \\
%     \textsuperscript{3}IBM Research Lab, Haifa, Israel\\
%     Email: lirans@il.ibm.com
% }

\author{
    Hasibul Jamil\textsuperscript{1,2*}, Abdul Alim\textsuperscript{1}, Laurent Schares\textsuperscript{1}, Pavlos Maniotis\textsuperscript{1},\\
    Liran Schour\textsuperscript{3}, Ali Sydney\textsuperscript{1}, Abdullah Kayi\textsuperscript{1}, Tevfik Kosar\textsuperscript{2}, Bengi Karacali\textsuperscript{1}\\
    \textsuperscript{1}IBM Research, Yorktown Heights, NY, USA. \\
    \textsuperscript{2}University at Buffalo (SUNY), Buffalo, NY, USA. \\
    \textsuperscript{3}IBM Research Lab, Haifa, Israel.\\
    Email: bkaraca@ibm.com\\
    \textsuperscript{*}This work was performed during an internship at IBM Research, Yorktown Heights, NY, USA.
}
\maketitle

\begin{abstract}
The increasing complexity of AI workloads, especially distributed Large Language Model (LLM) training, places significant strain on the networking infrastructure of parallel data centers and supercomputing systems. While Equal-Cost Multi-Path (ECMP) routing distributes traffic over parallel paths, hash collisions often lead to imbalanced network resource utilization and performance bottlenecks. This paper presents FlowTracer, a tool designed to analyze network path utilization and evaluate different routing strategies. FlowTracer aids in debugging network inefficiencies by providing detailed visibility into traffic distribution and helping to identify the root causes of performance degradation, such as issues caused by hash collisions. By offering flow-level insights, FlowTracer enables system operators to optimize routing, reduce congestion, and improve the performance of distributed AI workloads. We use a RoCEv2-enabled cluster with a leaf-spine network and 16 400-Gbps nodes to demonstrate how FlowTracer can be used to compare the flow imbalances of ECMP routing against a statically configured network. The example showcases a 30\% reduction in imbalance, as measured by a new metric we introduce.

\end{abstract}


\section{Introduction}
% \vspace{-1mm}
Recent advancements in artificial intelligence (AI), particularly deep learning and large-scale model training, have led to massive, distributed GPU-based workloads \cite{stellaTrain, crux, zero_infinity}. While GPUs provide essential computational power, successful training heavily depends on the underlying network infrastructure \cite{hedara,meta, rail-only, hpn, ibminfra}. Communication in AI training workloads is often realized via remote direct memory access (RDMA) traffic generated by multiple GPUs or other accelerators. On Ethernet fabrics, which are a popular solution for interconnecting AI servers, the most common protocol for this purpose is RDMA over Converged Ethernet (RoCE). The inter-GPU traffic typically consists of long-lived, high-bandwidth “elephant” flows. Since RDMA flows bypass the kernel, traditional network monitoring tools like tcpdump~\cite{tcpdump} are ineffective for analyzing this traffic. Data centers commonly rely on ECMP \cite{ecmp} routing to balance traffic across redundant paths, but this can result in ECMP hash collisions \cite{hedara}. These collisions cause network path imbalances, overloading some paths while underutilizing others, thereby degrading performance. RoCE flows, which can sustain hundreds of Gbps in a single flow, are particularly susceptible to ECMP collisions. For instance, the collision of two 100 Gbps RoCE flows at a 100 Gbps switch port results in a 100 Gbps loss or a 50 Gbps loss per flow. For applications with elephant flows, which are prone to such hash collisions, this leads to reduced performance and increased tail latencies.

This paper introduces FlowTracer, a tool designed to detect and quantify network imbalances by providing fine-grained visibility into the distribution of flows across physical network interfaces and links. FlowTracer is implemented in Python and can be used to identify issues such as ECMP collisions and spine-crossing inefficiencies, helping system operators debug performance degradation and pinpoint the root causes of imbalances. With runtimes as low as tens of seconds, it delivers detailed feedback by analyzing traffic from both servers and switches, enabling operators to optimize routing, reduce congestion, and improve resource utilization for enhanced AI workload performance. The key contributions of this paper are as follows:
\begin{itemize} 
    \item The design of FlowTracer and the parallel processing algorithm behind it. 
    \item The introduction of a new metric called Flow Imbalance Metric (FIM), which is crucial for comparing different network configurations and assessing their effectiveness. 
    \item A scalability analysis of the tool using TCP traffic, providing insights into efficiently configuring FlowTracer. 
    \item A detailed use case scenario where two different network configurations are examined using a RoCEv2-enabled cluster with a leaf-spine network and 16 400-Gbps nodes. 
\end{itemize}
 The paper is organized as follows: Section II covers background and related work, Section III details the FlowTracer design and architecture, Section IV presents evaluation and results, and Section V discusses future work and concludes the paper.

\vspace{-2mm}
\section{Background and Related Work}
\label{sec:background_related}
\vspace{-1mm}
Modern data centers supporting distributed AI workloads typically use folded Clos topologies \cite{hedara}, \cite{meta}, \cite{hpn}, \cite{ibminfra}, which offer multiple parallel paths between nodes for improved bandwidth and fault tolerance. ECMP routing, commonly used in Ethernet fabrics, hashes network flows and distributes them across multiple equal-cost paths. However, ECMP may lead to significant performance degradation when multiple flows are hashed to the same path. Additional challenges may also arise when interacting with overlay technologies like Virtual Extensible LAN (VXLAN) \cite{vxlan}. VXLAN encapsulates traffic to improve scalability and security when multiple virtual machines (VMs) are hosted on the same node. However, this encapsulation reduces the number of fields available for ECMP hash calculations, which limits overall entropy and makes hash collisions more likely \cite{googleHash}. This issue is particularly problematic for AI workloads generating large volumes of RoCE traffic, as imbalances in network utilization can severely degrade training performance.

Synchronization delays caused by these imbalances in distributed AI training can severely impact performance, as GPUs must synchronize at the end of each training iteration \cite{zero_infinity}. Uneven network utilization leads to bandwidth bottlenecks, reducing throughput and efficiency. To address this issue, the identification and quantification of network imbalances is critical.

Several tools exist for network health monitoring, but they often lack the detailed insights required for optimizing AI traffic patterns. Nagios \cite{nagios}, used for service monitoring, provides system health and availability overviews but lacks the granularity needed to trace specific network path utilization. Wireshark \cite{wireshark}, while effective for packet-level inspection, does not allow for analyzing how traffic is distributed across multiple network paths. Tools like Paris Traceroute \cite{parisTraceroute} and Dublin Traceroute \cite{dublinTraceroute} can map potential paths between two nodes, useful for detecting ECMP path diversity. However, they generate their own traffic and cannot trace the actual paths taken by other workloads, limiting their effectiveness in diagnosing flow imbalances.

More advanced tools like uMon \cite{umon} and Zoom2Net \cite{zoom2net} provide fine-grained insights but come with limitations. uMon offers microsecond-level granularity for capturing flow rate fluctuations and congestion. However, it introduces additional infrastructure requirements and overhead. Zoom2Net uses time-series imputation to enhance coarse-grained monitoring but lacks flow-specific insights. Active probing tools like R-Pingmesh \cite{rpingmesh} and RD-Probe \cite{rdprobe} diagnose RoCE traffic issues but introduce synthetic traffic into the network. FlowTracer distinguishes itself by passively monitoring real traffic flows, providing flow-level insights without injecting synthetic traffic.
FlowTracer uniquely combines the monitoring of workload-generated traffic with the detection of network path load imbalances, making it a valuable tool for debugging and optimizing network configurations in demanding environments such as AI systems.

\begin{figure}[!tbp]
  \centering
\includegraphics[width=\columnwidth]{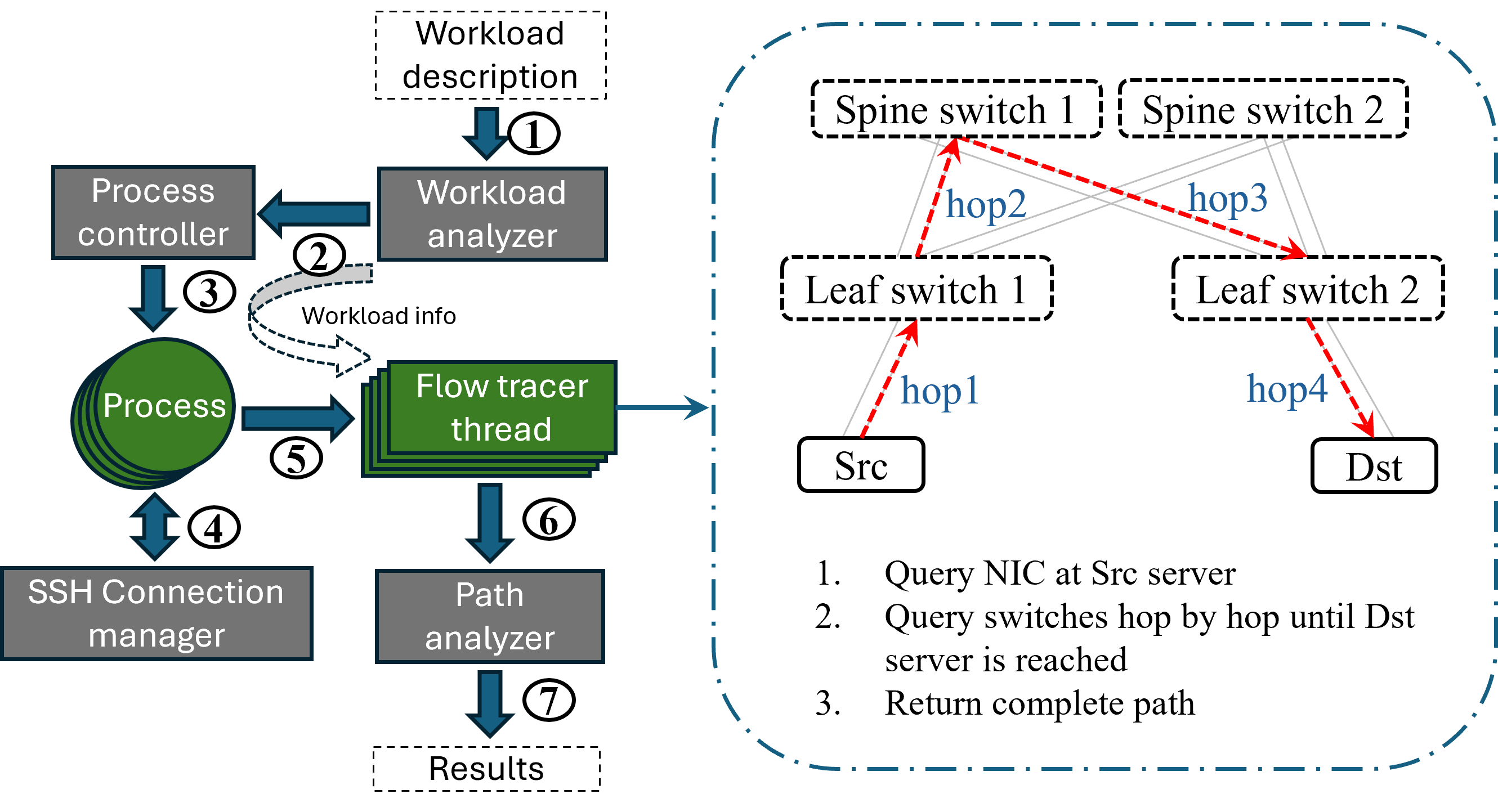}
  \caption{\textbf{FlowTracer's architecture and the process of hop-by-hop path discovery for all flows.}}
    \label{fig:systemBlock}
\vspace{-1mm}
\end{figure}

\vspace{-1mm}
\section{Design and Architecture of FlowTracer}
\label{sec:methodology}
% \vspace{-1mm}

FlowTracer is designed to trace network flows hop-by-hop for every flow originating from a specified set of servers, providing detailed visibility into network path utilization. FlowTracer is implemented in Python. A detailed block diagram of FlowTracer's architecture is illustrated in Figure~\ref{fig:systemBlock}, and the algorithm behind its operation is presented in Algorithm~\ref{alg:flowpath}

\subsection{Operation of FlowTracer}

FlowTracer takes as input the workload description (Step~\textcircled{1}), specifying the exact server pairs involved in the communication during the workload execution and the number of flows $f$ between these pairs. The \textit{Workload Analyzer} processes this information and triggers the \textit{Process Controller} (Step~\textcircled{2}), which is responsible for creating parallel processes to monitor one or more source-destination communication pairs. The number of processes can be set by the user or can be automatically calculated based on the total number of pairs in the workload. Next, the \textit{SSH Connection Manager} (Step~\textcircled{4}) is triggered to create the necessary SSH channels to exchange flow information with the corresponding servers and switches. Each process first extracts the flow’s 5-tuple information, which includes the source IP, destination IP, source port, destination port, and protocol. This applies to all flows originating from the source node of each communication pair handled by the process. FlowTracer ensures that only traffic related to the workload under test is analyzed, excluding any irrelevant traffic from the tracing process. In Step~\textcircled{5}, each process creates one or more \textit{FlowTracer threads}, further parallelizing the path discovery process, assigning each flow to a separate thread. As shown on the right side of the figure, each thread is responsible for discovering the flow's complete path through the topology on a hop-by-hop basis. For each hop, the thread identifies the outgoing interface of the flow by considering the ingress interface of the previous hop, i.e., it discovers the routing decision made for that flow by each device. A network topology file, which includes information about links and device interfaces, facilitates this process. Once all paths are discovered, the \textit{Path Analyzer} (Step~\textcircled{6}) is triggered to compile the final output, presenting the results in an easy-to-consume format (Step~\textcircled{7}).

Algorithm~\ref{alg:flowpath} outlines the parallel algorithm implemented by FlowTracer. The input includes workload information, the number of flows $f$ for each source-destination pair $(s, d)$, and an optional user-defined parameter that determines the number of parallel processes $P$ and the number of FlowTracer threads $T$ per process. Each $(s, d)$ pair is assigned to a process $P$, which establishes a communication channel with the source host $s$ to retrieve the 5-tuple information $\tau_{s,d}$ for all flows. Only flows relevant to the target workload are monitored. These flows are then divided into $T$ subsets, with each subset handled by a corresponding FlowTracer thread. Each thread is responsible for discovering the end-to-end path for its assigned flows. Once all flows are processed, the results are stored and returned as the final output.

% \setlength{\textfloatsep}{0pt}% Remove \textfloatsep

% \begin{algorithm}
% \caption{Algorithm behind the FlowTracer implementation}\label{alg:flowpath}
% \begin{algorithmic}[1]
% \STATE \textbf{Input:} Workload info: Source-destination pairs $(s, d)$ and number of flows $f$ per pair.
% User input: Number of parallel threads $P$.
% \STATE \textbf{Output:} End-to-end flow paths for all source-destination pairs
% \FOR{each source-destination pair $(s, d)$ }
%     \STATE Create a connection to the source node $s$.
%     \STATE Retrieve flow 5-tuple information $\tau_{s,d}$ for the $f$ flows between $s$ and $d$.
%     \STATE Filter $\tau_{s,d}$ to include only flows relevant to the current workload.
%     \STATE Divide the flows $\tau_{s,d}$ into $P$ subsets:
%     \vspace{-2mm}
%     \[
%     \{\tau_{s,d}^{(1)}, \tau_{s,d}^{(2)}, \dots, \tau_{s,d}^{(P)}\}, \quad \text{where} \quad \bigcup_{i=1}^{P} \tau_{s,d}^{(i)} = \tau_{s,d}
%     \]
%     \FOR{each thread $p =$ $1$ $to$ $P$ \textbf{in parallel}}
%         \FOR{each flow $\tau$ in $\tau_{s,d}^{(p)}$}
%             \STATE Trace the end-to-end path for $\tau$ using hop-by-hop discovery.
%         \ENDFOR
%     \ENDFOR
% \ENDFOR
% \STATE Store results for all source-destination pairs.
% \STATE \textbf{Return} the end-to-end flow paths for all source-destination pairs.
% \end{algorithmic}
% \end{algorithm}
\setlength{\textfloatsep}{0pt}% Remove \textfloatsep

\begin{algorithm}[t]
\caption{FlowTracer Parallel Path Discovery Algorithm}\label{alg:flowpath}
\begin{algorithmic}[1]
\raggedright
\STATE \textbf{Input:} Workload information: Source-destination pairs $(s, d)$, number of flows $f$ per pair, filter info. \\
User-defined input: Number of parallel processes $P$, number of threads $T$ per process.
\STATE \textbf{Output:} End-to-end flow paths for all $(s,d)$ pairs.
\STATE Divide $(s, d)$ pairs among the processes $P$. These processes are executed concurrently.
\FOR{each $(s, d)$ pair assigned to process $P$}
    \STATE Establish communication channel with $s$.
    \STATE Retrieve flow 5-tuple information $\tau_{s,d}$ for the $f$ flows between $s$ and $d$.
    \STATE Filter $\tau_{s,d}$ to include only flows relevant to the current workload based on filter info.
    \STATE Divide the flows $\tau_{s,d}$ into $T$ subsets for parallel processing:
    \vspace{-2mm}
    \[
    \{\tau_{s,d}^{(1)}, \tau_{s,d}^{(2)}, \dots, \tau_{s,d}^{(T)}\}, \quad \text{where} \quad \bigcup_{i=1}^{T} \tau_{s,d}^{(i)} = \tau_{s,d}
    \]
    \FOR{each FlowTracer thread $t = 1$ to $T$ \textbf{in parallel}}
        \FOR{each flow $\tau$ in $\tau_{s,d}^{(t)}$}
            \STATE Trace the end-to-end path for $\tau$ using hop-by-hop path discovery.
        \ENDFOR
    \ENDFOR
\ENDFOR
\STATE Store results for all $(s, d)$ pairs.
\STATE \textbf{Return} the end-to-end flow paths for all $(s, d)$ pairs.
\end{algorithmic}
\end{algorithm}
% \end{minipage}

% \vspace{-5mm}
\subsection{Hop-by-Hop path discovery}
% \vspace{-1mm}
FlowTracer tracks the network flows hop-by-hop to provide end-to-end path visibility. The process begins by identifying the first hop, which varies depending on whether the traffic is handled by the operating system's kernel (e.g., TCP/UDP) or bypasses it (e.g., RDMA traffic). After identifying the first hop, FlowTracer continues tracing the flow through each subsequent hop until the destination is reached. Below, we describe the process for both TCP/UDP and RDMA traffic, followed by a detailed explanation of the hop-by-hop discovery process.
\subsubsection{\textbf{First Hop Tracking}}
\paragraph{TCP/UDP traffic}

For TCP/UDP traffic, FlowTracer uses kernel utilities (such as the \texttt{ss} tool~\cite{ss}) and the routing table on the source node to retrieve the flow’s 5-tuple information and determine the outgoing interface. After retrieving the flow tuple, FlowTracer uses the sender’s kernel routing table to identify the physical egress interface for the first hop. This ensures that the flow is correctly routed from the source node into the network and directed toward the first network entity, typically a leaf switch.

\paragraph{RDMA traffic}

Since RDMA traffic bypasses the traditional operating system kernel's network stack, FlowTracer interfaces directly with the NIC driver to retrieve the 5-tuple information for each flow. The NIC driver plays a key role in logging flow details, as the kernel routing table is not applicable for this type of traffic. For identifying the physical egress interface of the first hop, FlowTracer supports both a NIC driver-based approach and an sFlow-based approach \cite{sflow}. The NIC driver approach retrieves interface usage directly from the network interface card. However, to address potential network driver support issues, we also implemented an sFlow-based method, where we collect ongoing flow information from leaf switches only.

\subsubsection{\textbf{Subsequent Hop-by-Hop Tracking}}

Once the first hop has been identified, FlowTracer proceeds with hop-by-hop tracking for the remainder of the flow's journey through the network. This process is the same for both TCP/UDP and RDMA traffic. FlowTracer queries each network switch along the path to determine the outgoing interface for the flow. To achieve that, we used the ECMP hash visibility CLI from Arista \cite{arista-cli}. At each hop, FlowTracer relies on the 5-tuple information and the ingress interface from the previous hop to compute the outgoing interface. The ingress interface is derived from the egress interface of the previous network entity using the network topology file, as described in Section III.A. This process is repeated at each network switch—starting from the leaf switch, followed by the spine switches, and then back to the leaf switches—until the flow reaches its final destination.

\subsection{Flow Imbalance Metric}

To quantify the load distribution across network interface links, we define the \textit{Flow Imbalance Metric} (FIM). This metric measures the deviation of actual flow counts from the ideal distribution. The ideal scenario occurs when each network link at each level of the network carries an equal number of flows, resulting in a balanced load and optimized throughput. 
The FIM is computed using the Mean Absolute Percentage Error (MAPE) as follows:

\begin{equation}
    \text{FIM} = \frac{1}{n} \sum_{i=1}^{n} \left| \frac{\text{actual\_flows}_i - \text{ideal\_flows}_i}{\text{ideal\_flows}_i} \right| \times 100
\end{equation}

where $n$ is the total number of network links available in the fabric, $\text{actual\_flows}_i$ is the number of flows on the $i$-th link, and $\text{ideal\_flows}_i$ is the ideal number of flows per link in a perfectly balanced scenario. A lower FIM value indicates a more balanced distribution of traffic across the network, leading to better throughput performance. On the other hand, a higher FIM suggests an uneven load distribution, which can result in bottlenecks and suboptimal network performance.

% \vspace{-2mm}
\section{Evaluation and Analysis}
\label{sec:evaluation}
\begin{figure}
  \centering
\includegraphics[width=0.8\columnwidth]{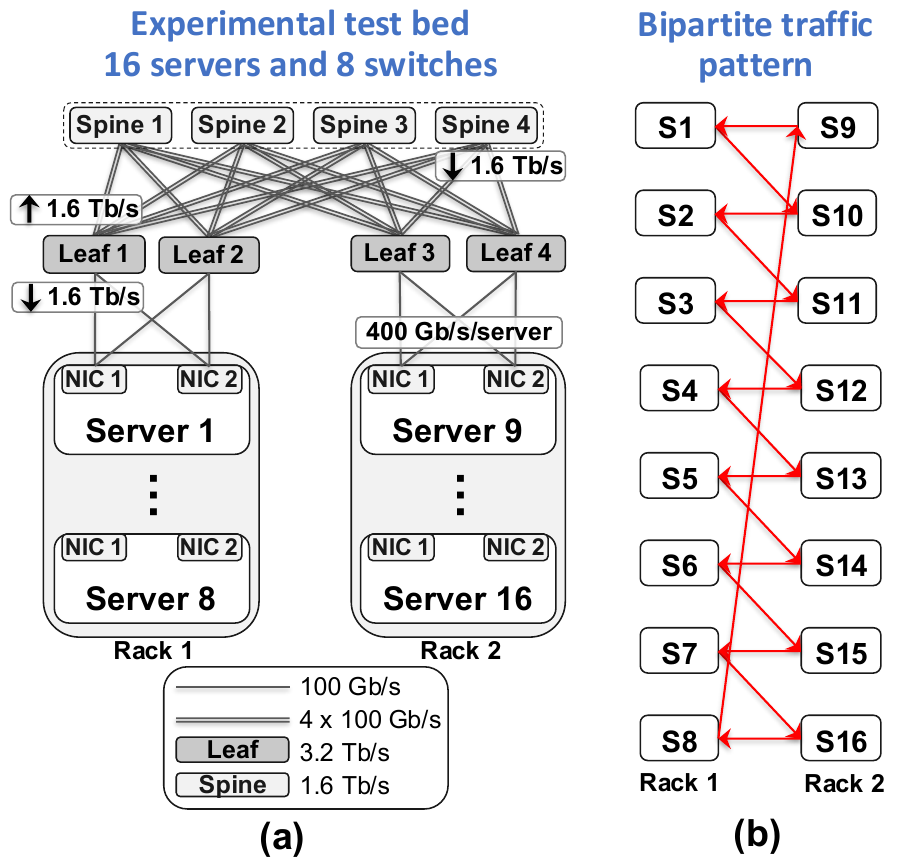}
  \caption{\textbf{(a) Experimental 2-rack testbed consisting of 16 servers and 8 switches. Each server has two dual-port 100-Gb/s NICs, i.e., a total bandwidth of 400 Gb/s per server. The network contains four spine and four leaf switches, with a 1.6-Tb/s cross-rack bandwidth between the leaf and spine layers. (b): bipartite traffic pattern between servers in the two racks that are used in our evaluation.}}
  \vspace{2mm}
    \label{fig:network_tree}
\end{figure}

FlowTracer is a light-weight tool and can run on any physical or virtual machine, provided it can connect to the target servers and switches. Figure~\ref{fig:network_tree}(a) depicts the testbed used for the experiments, following modern data center principles. The network operates with a 1:1 subscription ratio, utilizing 1.6 Tbps spine switches and 3.2 Tbps leaf switches, with all links running at 100 Gbps. The testbed includes 16 servers with 56-core Intel Xeon CPUs running Ubuntu 22.04, 540 GiB of RAM, NVIDIA CX6-DX NICs, and with an in-house developed software switch that enables server-based virtualization using VXLAN encapsulation. Each server connects to two Top-of-Rack switches, and VMs with SR-IOV-enabled NVIDIA OFED (MLNX\_OFED) 5.4 are deployed, supporting RoCEv2 traffic with adaptive retransmission. Inter-rack communication spans 3 hops, while intra-rack communication requires 1 hop.

Figure ~\ref{fig:network_tree}(b) illustrates the bipartite traffic pattern used for our tests, which generates sufficient cross-rack traffic to saturate 100\% of the links between the two racks. This traffic pattern was selected to stress cross-rack connectivity, which is often observed in distributed AI training workloads. Since there are four ECMP hashing decisions involved, the likelihood of hash collisions is higher compared to intra-rack traffic, which involves only two ECMP decisions.

AI workloads typically utilize all-reduce collectives that operate by creating logical rings. These logical rings can be constructed in various ways, with different network footprints. For example, when most of the edges of the logical rings are contained within the rack boundaries of the cluster, the bulk of the traffic traverses a single level of network switches, thereby reducing the probability of ECMP hashing collisions. In the worst case, however, all ring edges could map to network paths that cross the spine switches, effectively forming a bipartite graph. This, in turn, increases the probability of ECMP hashing collisions, which can significantly impact workload performance.

\subsection{A Use Case Analysis: Evaluating Routing Configurations With RoCEv2 Traffic}

In this analysis, we generate RoCE traffic using NVIDIA’s perftest package \cite{ibperftest}, specifically \texttt{ib\_send\_bw} and \texttt{ib\_write\_bw}, and we create multiple source-destination connections according to the bipartite traffic pattern depicted in Figure ~\ref{fig:network_tree}(b), i.e., 256 RoCE flows in total, which corresponds to 4 number of flows per link for a perfectly balanced distribution. All our tests are repeated multiple times for statistical significance.  

\begin{figure*}[!htbp]
    \centering
    \includegraphics[width=0.94\textwidth]{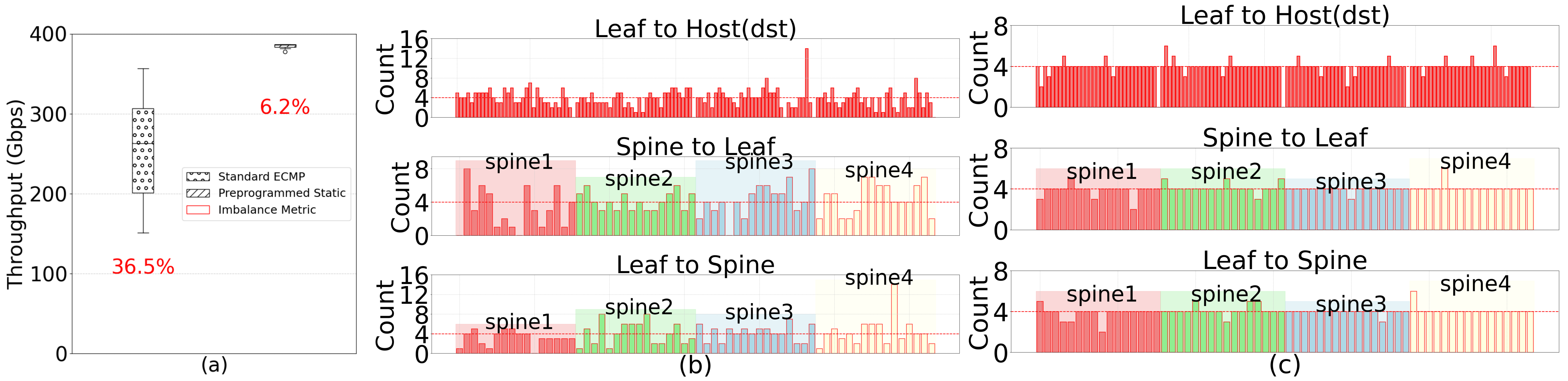}
    \caption{\textbf{(a) RoCE throughput distribution across all node pairs and their corresponding imbalance metric shown in red (lower is better). (b) flow distributions of 256 RoCE flows for standard ECMP-based routing. (c) flow distributions for preprogrammed static routing. The red line in each subfigure represents the ideal flow distribution, i.e., all flows are balanced perfectly. Standard ECMP routing results in a noticeable load imbalance, while static routing provides much more balanced distributions.}}
    \label{fig:combined}
    \vspace{-6mm}
\end{figure*}

Figure ~\ref{fig:combined}(a) shows the throughput distribution for all 16 server pairs. The analysis is performed for the following two network routing configurations: (a) a standard ECMP-based routing configuration (left boxplot), and (b) a preprogrammed static routing configuration (right boxplot), which promotes the selection of distinct paths across the different communication pairs, thereby avoiding flow collisions. Despite the network topology offering a 1:1 oversubscription ratio and full bisection bandwidth, we observe a significant performance impact in the case of standard ECMP routing. This is the result of ECMP hash collisions, which cause multiple flows to be mapped to the same network paths, leading to congestion and underutilization of other available paths. On the other hand, we observe that the static routing configuration exhibits near-line-rate performance for all servers, which is also confirmed by the Flow Imbalance Metric (shown in red), as it is in agreement with the corresponding boxplots. Specifically, the more spread-out distribution in the standard ECMP case shows a higher imbalance metric of 36.5\%, while the less spread-out distribution in the static routing case exhibits a lower imbalance metric of 6.2\%.

Figures ~\ref{fig:combined}(b) and ~\ref{fig:combined}(c) present the detailed flow distributions across the network links for both routing configurations, covering all the leaf-to-spine, spine-to-leaf, and leaf-to-host link layers. The red horizontal lines in each subfigure represent the ideal scenario, where flows are perfectly distributed evenly across all network links. Specifically, Figure ~\ref{fig:combined}(b) illustrates the flow distribution for the standard ECMP routing case, where hash collisions result in a highly uneven distribution at all layers. In contrast, Figure ~\ref{fig:combined}(c) depicts the flow distribution for the static routing case, where the distributions are much more balanced and closely aligned with the ideal case.

\subsection{Scalability Analysis: Evaluating FlowTracer's Performance With TCP traffic}

\begin{figure} [t]
  \centering
\includegraphics[width=\columnwidth]{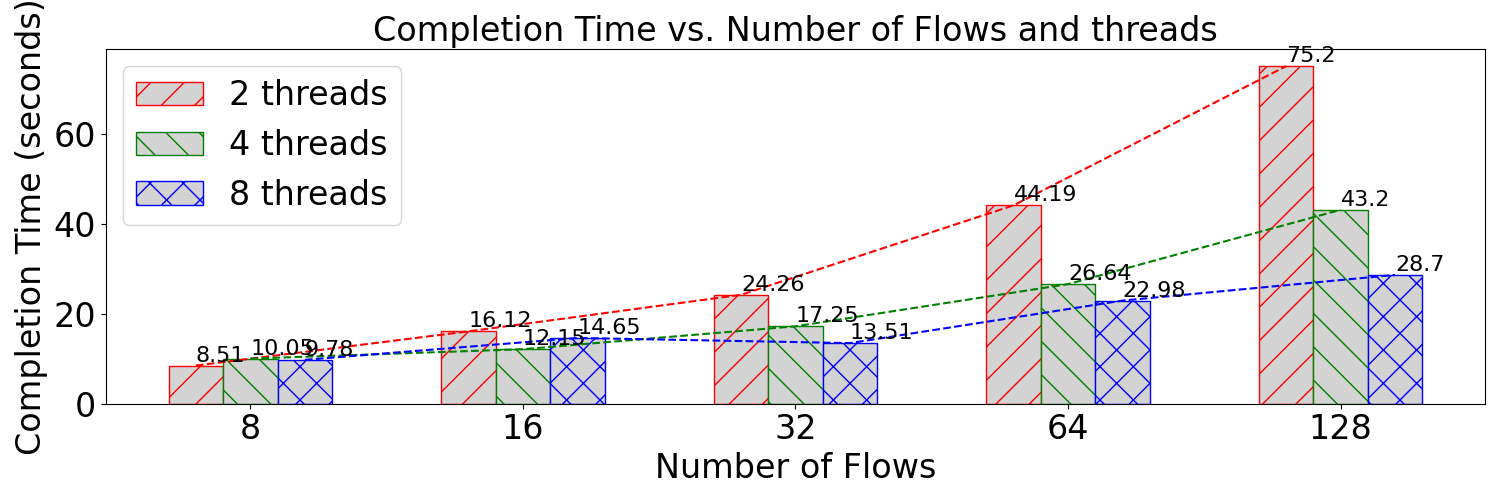}
  \caption{\textbf{Comparison of completion time versus the number of flows for different numbers of parallel threads (2, 4, and 8 threads).}}
\vspace{3mm}
\label{fig:scalabilityWithThreads}
\end{figure}

\begin{figure}[t]
  \centering
\includegraphics[width=\columnwidth]{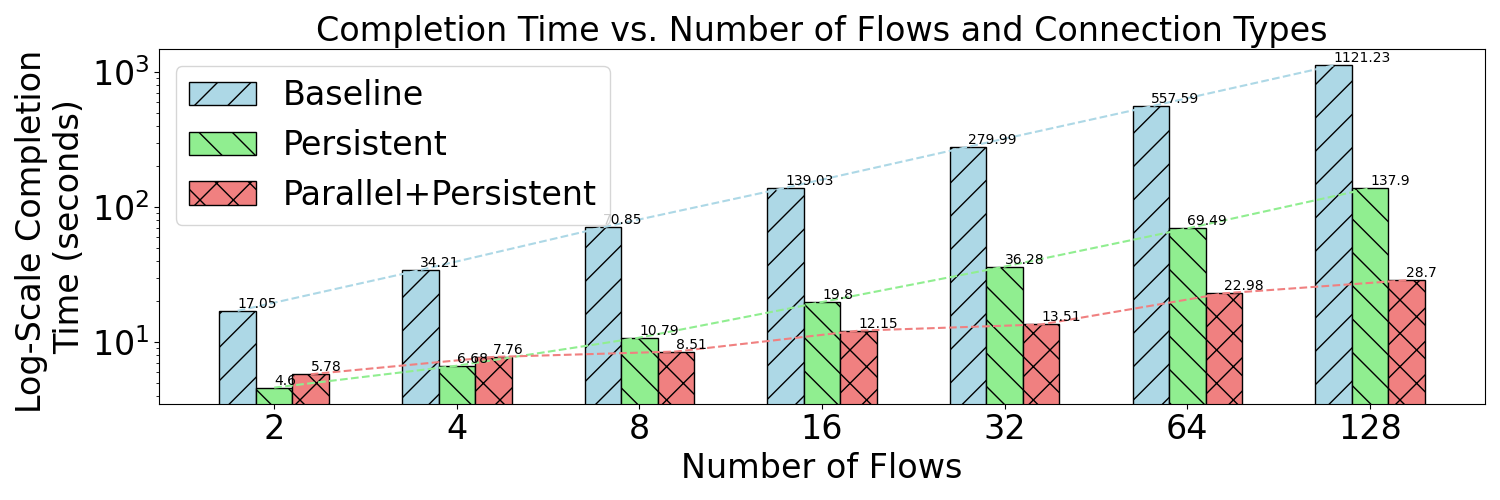}
  \caption{\textbf{Completion time versus number of flows for three different SSH connection approaches with remote devices: Baseline, Persistent, and Parallel+Persistent. The Baseline configuration represents ad-hoc SSH connections that are immediately terminated after the path discovery. The Persistent configuration uses a single SSH connection for multiple path discoveries, leading to reduced completion times compared to Baseline. The Parallel+Persistent configuration leverages multiple persistent SSH connections managed by parallel threads, resulting in the fastest completion times.}}
\vspace{3mm}
\label{fig:connectionTypes}
\end{figure}

Figures ~\ref{fig:scalabilityWithThreads} and ~\ref{fig:connectionTypes} present FlowTracer's scalability properties by varying the number of flows and parallel processes/threads, as well as by examining different methods of establishing SSH connections with remote devices (i.e., servers and switches). As expected, FlowTracer's overall runtime is inversely proportional to the number of parallel processes/threads used in path discovery.

In this analysis, we generate TCP traffic using iPerf3 \cite{iperf3}, specifically employing parallel TCP streams (the terms streams and flows are used interchangeably in this case). The scalability analysis was performed for a subset of source-destination pairs of the traffic pattern shown in Figure ~\ref{fig:network_tree}(b). Figures ~\ref{fig:scalabilityWithThreads} and ~\ref{fig:connectionTypes} present the average measurements collected from all the source-destination pairs.

Figure ~\ref{fig:scalabilityWithThreads} focuses on the runtime for path discovery for a set of number of flows and number of threads. As the number of flows increases, the completion time increases proportionally for all thread configurations (2, 4, and 8 threads); however, systems utilizing more parallel threads achieve shorter completion times. This highlights FlowTracer's ability to efficiently take advantage of parallelism when tracing large numbers of flows. For instance, the completion time for 128 flows is reduced by a factor of 2.6 when using 8 parallel threads compared to 2.

Figure ~\ref{fig:connectionTypes} compares three different approaches for establishing SSH connections with the remote devices (i.e., the servers and the switches): Baseline, Persistent, and Parallel+Persistent. The Baseline configuration uses one thread and ad-hoc SSH connections that are immediately closed after each flow is traced, resulting in higher overhead. In contrast, the Persistent configuration, which still uses one thread, maintains a single SSH connection for tracing multiple flows and only closes the connection after all flows have been traced, thereby reducing overhead. The Parallel+Persistent configuration combines persistent SSH connections with multiple parallel threads—8 threads for the cases of 8 or more flows, and 2 and 4 threads for the cases of 2 and 4 flows, respectively—allowing for concurrent flow tracing while maintaining persistent connections. This configuration demonstrates the best scalability, achieving the lowest completion times as the number of flows increases.

As networks scale and the number of switches and servers increases, tracing the flows naturally takes longer due to the larger infrastructure. To address this, FlowTracer can scale by partitioning the system into subsystems and running parallel instances of the tool for each partition, while periodic synchronization can be used to exchange information between partitions. This approach distributes the workload among the parallel FlowTracer instances, keeping tracing times manageable. Additionally, the current use of SSH for communication introduces overhead, which could be reduced by adopting a customized, more efficient protocol. These strategies would enable FlowTracer to scale effectively for larger networks and complex distributed AI workloads.

\vspace{-2mm}
\section{Conclusion and Future Work}
\label{sec:conclusion}
\vspace{-1mm}

In this paper, we presented FlowTracer, a tool designed to address network path imbalances in distributed AI workloads, particularly in the context of LLM training. FlowTracer provides detailed visibility into traffic distribution, helping to identify the root causes of performance degradation, such as issues caused by ECMP hash collisions. It can be used as a debugging tool to enhance network resource utilization and improve the overall performance of distributed AI workloads. In a use case example, we demonstrated how FlowTracer can be used to compare different network configurations and quantify link load imbalance as measured by a newly proposed metric introduced in this paper.

For future work, we aim to enhance FlowTracer with predictive models that leverage real-time flow data, enabling real-time monitoring of individual flow characteristics and dynamic routing adjustments. These improvements will facilitate both proactive and reactive optimization techniques, maximizing performance and scalability in AI workloads.

\bibliographystyle{IEEEtran}
% \bibliography{references}
\bibliography{reference}

\begin{thebibliography}{10}
\providecommand{\url}[1]{#1}
\csname url@samestyle\endcsname
\providecommand{\newblock}{\relax}
\providecommand{\bibinfo}[2]{#2}
\providecommand{\BIBentrySTDinterwordspacing}{\spaceskip=0pt\relax}
\providecommand{\BIBentryALTinterwordstretchfactor}{4}
\providecommand{\BIBentryALTinterwordspacing}{\spaceskip=\fontdimen2\font plus
\BIBentryALTinterwordstretchfactor\fontdimen3\font minus \fontdimen4\font\relax}
\providecommand{\BIBforeignlanguage}[2]{{%
\expandafter\ifx\csname l@#1\endcsname\relax
\typeout{** WARNING: IEEEtran.bst: No hyphenation pattern has been}%
\typeout{** loaded for the language `#1'. Using the pattern for}%
\typeout{** the default language instead.}%
\else
\language=\csname l@#1\endcsname
\fi
#2}}
\providecommand{\BIBdecl}{\relax}
\BIBdecl

\bibitem{stellaTrain}
\BIBentryALTinterwordspacing
H.~Lim, J.~Ye, S.~Abdu~Jyothi, and D.~Han, ``Accelerating model training in multi-cluster environments with consumer-grade gpus,'' in \emph{Proceedings of the ACM SIGCOMM 2024 Conference}, ser. ACM SIGCOMM '24.\hskip 1em plus 0.5em minus 0.4em\relax New York, NY, USA: Association for Computing Machinery, 2024, p. 707–720. [Online]. Available: \url{https://doi.org/10.1145/3651890.3672228}
\BIBentrySTDinterwordspacing

\bibitem{crux}
\BIBentryALTinterwordspacing
J.~Cao, Y.~Guan, K.~Qian, J.~Gao, W.~Xiao, J.~Dong, B.~Fu, D.~Cai, and E.~Zhai, ``Crux: Gpu-efficient communication scheduling for deep learning training,'' in \emph{Proceedings of the ACM SIGCOMM 2024 Conference}, ser. ACM SIGCOMM '24.\hskip 1em plus 0.5em minus 0.4em\relax New York, NY, USA: Association for Computing Machinery, 2024, p. 1–15. [Online]. Available: \url{https://doi.org/10.1145/3651890.3672239}
\BIBentrySTDinterwordspacing

\bibitem{zero_infinity}
\BIBentryALTinterwordspacing
S.~Rajbhandari, O.~Ruwase, J.~Rasley, S.~Smith, and Y.~He, ``Zero-infinity: breaking the gpu memory wall for extreme scale deep learning,'' in \emph{Proceedings of the International Conference for High Performance Computing, Networking, Storage and Analysis}, ser. SC '21.\hskip 1em plus 0.5em minus 0.4em\relax New York, NY, USA: Association for Computing Machinery, 2021. [Online]. Available: \url{https://doi.org/10.1145/3458817.3476205}
\BIBentrySTDinterwordspacing

\bibitem{hedara}
M.~Al-Fares, S.~Radhakrishnan, B.~Raghavan, N.~Huang, and A.~Vahdat, ``Hedera: dynamic flow scheduling for data center networks,'' in \emph{Proceedings of the 7th USENIX Conference on Networked Systems Design and Implementation}, ser. NSDI'10.\hskip 1em plus 0.5em minus 0.4em\relax USENIX Association, 2010, p.~19.

\bibitem{meta}
\BIBentryALTinterwordspacing
A.~Gangidi, R.~Miao, S.~Zheng, S.~J. Bondu, G.~Goes, H.~Morsy, R.~Puri, M.~Riftadi, A.~J. Shetty, J.~Yang, S.~Zhang, M.~J. Fernandez, S.~Gandham, and H.~Zeng, ``Rdma over ethernet for distributed training at meta scale,'' in \emph{Proceedings of the ACM SIGCOMM 2024 Conference}, ser. ACM SIGCOMM '24.\hskip 1em plus 0.5em minus 0.4em\relax New York, NY, USA: Association for Computing Machinery, 2024, p. 57–70. [Online]. Available: \url{https://doi.org/10.1145/3651890.3672233}
\BIBentrySTDinterwordspacing

\bibitem{rail-only}
\BIBentryALTinterwordspacing
W.~Wang, M.~Ghobadi, K.~Shakeri, Y.~Zhang, and N.~Hasani, ``Rail-only: A low-cost high-performance network for training llms with trillion parameters,'' 2024. [Online]. Available: \url{https://arxiv.org/abs/2307.12169}
\BIBentrySTDinterwordspacing

\bibitem{hpn}
\BIBentryALTinterwordspacing
K.~Qian, Y.~Xi, J.~Cao, J.~Gao, Y.~Xu, Y.~Guan, B.~Fu, X.~Shi, F.~Zhu, R.~Miao, C.~Wang, P.~Wang, P.~Zhang, X.~Zeng, E.~Ruan, Z.~Yao, E.~Zhai, and D.~Cai, ``Alibaba hpn: A data center network for large language model training,'' in \emph{Proceedings of the ACM SIGCOMM 2024 Conference}, ser. ACM SIGCOMM '24.\hskip 1em plus 0.5em minus 0.4em\relax New York, NY, USA: Association for Computing Machinery, 2024, p. 691–706. [Online]. Available: \url{https://doi.org/10.1145/3651890.3672265}
\BIBentrySTDinterwordspacing

\bibitem{ibminfra}
\BIBentryALTinterwordspacing
T.~G. et~al., ``The infrastructure powering ibm's gen ai model development,'' 2024. [Online]. Available: \url{https://arxiv.org/abs/2407.05467}
\BIBentrySTDinterwordspacing

\bibitem{tcpdump}
``tcpdump \& libpcap,'' \url{https://www.tcpdump.org}, accessed: 2024-10-09.

\bibitem{ecmp}
C.~Hopps, ``Rfc2992: Analysis of an equal-cost multi-path algorithm,'' USA, 2000.

\bibitem{vxlan}
\BIBentryALTinterwordspacing
M.~Mahalingam, D.~Dutt, K.~Duda, P.~Agarwal, L.~Kreeger, T.~Sridhar, M.~Bursell, and C.~Wright, ``{Virtual eXtensible Local Area Network (VXLAN): A Framework for Overlaying Virtualized Layer 2 Networks over Layer 3 Networks},'' RFC 7348, Aug. 2014. [Online]. Available: \url{https://www.rfc-editor.org/info/rfc7348}
\BIBentrySTDinterwordspacing

\bibitem{googleHash}
\BIBentryALTinterwordspacing
Y.~Xu, K.~He, R.~Wang, M.~Yu, N.~Duffield, H.~Wassel, S.~Zhang, L.~Poutievski, J.~Zhou, and A.~Vahdat, ``Hashing design in modern networks: Challenges and mitigation techniques,'' in \emph{2022 USENIX Annual Technical Conference (USENIX ATC 22)}.\hskip 1em plus 0.5em minus 0.4em\relax Carlsbad, CA: USENIX Association, Jul. 2022, pp. 805--818. [Online]. Available: \url{https://www.usenix.org/conference/atc22/presentation/xu}
\BIBentrySTDinterwordspacing

\bibitem{nagios}
``{Nagios},'' https://www.nagios.org/, 2024.

\bibitem{wireshark}
``{wireshark},'' https://www.wireshark.org/, 2024.

\bibitem{parisTraceroute}
\BIBentryALTinterwordspacing
B.~Augustin, T.~Friedman, and R.~Teixeira, ``Measuring load-balanced paths in the internet,'' in \emph{Proceedings of the 7th ACM SIGCOMM Conference on Internet Measurement}, ser. IMC '07.\hskip 1em plus 0.5em minus 0.4em\relax New York, NY, USA: Association for Computing Machinery, 2007, p. 149–160. [Online]. Available: \url{https://doi.org/10.1145/1298306.1298329}
\BIBentrySTDinterwordspacing

\bibitem{dublinTraceroute}
``{Dublin Traceroute},'' https://dublin-traceroute.net/, 2024.

\bibitem{umon}
\BIBentryALTinterwordspacing
H.~Zheng, C.~Huang, X.~Han, J.~Zheng, X.~Wang, C.~Tian, W.~Dou, and G.~Chen, ``umon: Empowering microsecond-level network monitoring with wavelets,'' in \emph{Proceedings of the ACM SIGCOMM 2024 Conference}, ser. ACM SIGCOMM '24.\hskip 1em plus 0.5em minus 0.4em\relax New York, NY, USA: Association for Computing Machinery, 2024, p. 274–290. [Online]. Available: \url{https://doi.org/10.1145/3651890.3672236}
\BIBentrySTDinterwordspacing

\bibitem{zoom2net}
\BIBentryALTinterwordspacing
F.~Gong, D.~Raghunathan, A.~Gupta, and M.~Apostolaki, ``Zoom2net: Constrained network telemetry imputation,'' in \emph{Proceedings of the ACM SIGCOMM 2024 Conference}, ser. ACM SIGCOMM '24.\hskip 1em plus 0.5em minus 0.4em\relax New York, NY, USA: Association for Computing Machinery, 2024, p. 764–777. [Online]. Available: \url{https://doi.org/10.1145/3651890.3672225}
\BIBentrySTDinterwordspacing

\bibitem{rpingmesh}
\BIBentryALTinterwordspacing
K.~Liu, Z.~Jiang, J.~Zhang, S.~Guo, X.~Zhang, Y.~Bai, Y.~Dong, F.~Luo, Z.~Zhang, L.~Wang, X.~Shi, H.~Xu, Y.~Bai, D.~Song, H.~Wei, B.~Li, Y.~Pan, T.~Pan, and T.~Huang, ``R-pingmesh: A service-aware roce network monitoring and diagnostic system,'' in \emph{Proceedings of the ACM SIGCOMM 2024 Conference}, ser. ACM SIGCOMM '24.\hskip 1em plus 0.5em minus 0.4em\relax New York, NY, USA: Association for Computing Machinery, 2024, p. 554–567. [Online]. Available: \url{https://doi.org/10.1145/3651890.3672264}
\BIBentrySTDinterwordspacing

\bibitem{rdprobe}
\BIBentryALTinterwordspacing
R.~Ding, X.~Liu, S.~Yang, Q.~Huang, B.~Xie, R.~Sun, Z.~Zhang, and B.~Cui, ``Rd-probe: Scalable monitoring with sufficient coverage in complex datacenter networks,'' in \emph{Proceedings of the ACM SIGCOMM 2024 Conference}, ser. ACM SIGCOMM '24.\hskip 1em plus 0.5em minus 0.4em\relax New York, NY, USA: Association for Computing Machinery, 2024, p. 258–273. [Online]. Available: \url{https://doi.org/10.1145/3651890.3672256}
\BIBentrySTDinterwordspacing

\bibitem{ss}
A.~Kuznetsov, ``{ss(8) - Linux manual page},'' \url{https://man7.org/linux/man-pages/man8/ss.8.html}, accessed: September 27, 2024.

\bibitem{sflow}
\BIBentryALTinterwordspacing
S.~Panchen, N.~McKee, and P.~Phaal, ``{InMon Corporation's sFlow: A Method for Monitoring Traffic in Switched and Routed Networks},'' RFC 3176, Sep. 2001. [Online]. Available: \url{https://www.rfc-editor.org/info/rfc3176}
\BIBentrySTDinterwordspacing

\bibitem{arista-cli}
``Ecmp hash visibility,'' \url{https://www.arista.com/en/support/toi/tag/hash-visibility }, accessed: 2024-10-09.

\bibitem{ibperftest}
``perftest,'' \url{https://enterprise-support.nvidia.com/s/article/perftest-package}, accessed: 2024-10-10.

\bibitem{iperf3}
``iperf3,'' \url{https://github.com/esnet/iperf }, accessed: 2024-10-09.

\end{thebibliography}

\end{document}